\documentclass{llncs}
\usepackage{llncsdoc}
\usepackage{times}
\usepackage{url}
\usepackage{cite}
\usepackage{amsmath}
\usepackage{amsfonts}
\usepackage{graphicx}
\usepackage{adjustbox}
\usepackage{subcaption}
\usepackage[ruled,linesnumbered]{algorithm2e}
\usepackage{url}
\usepackage{multirow}
\usepackage{float}
\newcommand{\R}{\mathbb{R}}

\begin{document}
\title{SNE: Signed Network Embedding}
\author{Shuhan Yuan \inst{1} \and Xintao Wu \inst{2} \and Yang Xiang \inst{1}}
\institute{Tongji University, Shanghai, China, Email:\email{\{4e66,shxiangyang\}@tongji.edu.cn}
\and University of Arkansas, Fayetteville, AR, USA, Email:\email{xintaowu@uark.edu}}

\maketitle
\begin{abstract}
Several network embedding models have been developed for unsigned networks. However, these models based on skip-gram cannot be applied to signed networks  because they can only deal with one type of link. In this paper, we present our signed network embedding model called SNE. Our SNE  adopts the log-bilinear model, uses node representations of all nodes along a given path, and further incorporates two signed-type vectors to capture the positive or negative relationship of each edge along the path. We conduct two experiments, node classification and link prediction, on both directed and undirected signed networks and compare with four  baselines including a matrix factorization method and three state-of-the-art unsigned network embedding models. The experimental results demonstrate the effectiveness of our signed network embedding.
\end{abstract}
\section{Introduction}

Representation learning \cite{Bengio2013Representation}, which aims to learn the features automatically based on various deep learning models \cite{Lecun2015Deep}, has been extensively studied in recent years.
Traditionally, supervised learning tasks require hand-designed features as inputs. Deep learning models have shown great success in automatically learning the semantic representations for different types of data, like image, text and speech \cite{Krizhevsky2012Imagenet,Collobert2011Natural,Graves2013Speech}. In this paper, we focus on representation learning of networks, in particular, signed networks.
Several representation learning methods of unsigned networks have been developed recently \cite{Perozzi2014Deepwalk,Tang2015Line,Grover2016Node2Vec,Tian2014Learning}.
They represent each node as a low-dimensional vector which  captures the structure information of the network.

Signed networks are ubiquitous in real-world social systems, which have both positive and negative relationships. For example, Epinions~\footnote{\url{http://www.epinions.com/}} allows users to mark their trust or distrust to other users on product reviews and Slashdot~\footnote{\url{https://slashdot.org/}} allows users to specify other users as friends or foes.  Most unsigned network embedding models \cite{Perozzi2014Deepwalk,Tang2015Line,Grover2016Node2Vec} are based on skip-gram \cite{Mikolov2013Efficient}, a classic approach for training word embeddings. The objective functions used in unsigned network embedding approaches  do not incorporate the sign information of edges. Thus, they cannot simply migrate to signed networks because the negative links change the theories or assumptions on which unsigned network embedding models rely \cite{Tang2015Survey}.

In this paper, we develop a signed network embedding model called SNE. To the best of our knowledge, this is the first research on  signed network embedding.
Our SNE model adopts the log-bilinear model \cite{Mnih2013Learning,Mnih2008Scalable}, uses node representations of all nodes along a given path, and further incorporates two signed-type vectors to capture the positive or negative relationship of each edge along the path.
Our SNE significantly outperforms existing unsigned network embedding models which assume  all edges are from the same type of relationship and only use the representations of nodes in the target's neighborhood.
We conduct two experiments to evaluate our model, node classification and link prediction, on both an undirected signed network and a directed signed network built from real-world data.
We compare with four  baselines including a matrix factorization method  and three state-of-the-art network embedding models designed for unsigned networks.
The experimental results demonstrate the effectiveness of our signed network embedding.

\vspace{-0.2cm}
\section{Preliminary}
In this section, we first introduce the skip-gram model, one of commonly used neural language models to train word embeddings \cite{Bengio2003Neural}. We then
give a brief overview of several state-of-the-art unsigned network embedding models based on the skip-gram model.

{\noindent \bf Skip-gram model}
The skip-gram is to model the co-occurrence probability $p(w_j|w_i;\theta)$ that word $w_j$ co-occurs with word $w_i$ in a context window. The co-occurrence probability is calculated based on softmax function:
\begin{equation}
\label{eq:softmax}
p(w_j|w_i;\theta)=\frac{\exp(\mathbf{v}_{w_j}^T \mathbf{v}_{w_i})}{\sum_{j' \in \mathcal{V}} \exp(\mathbf{v}_{w_{j'}}^T \mathbf{v}_{w_i})},
\end{equation}
where $\mathcal{V}$ is the set of all words and $\mathbf{v}_{w_j}$ and $\mathbf{v}_{w_i} \in \R^d$ are word embeddings for $w_j$ and $w_i$, respectively.  The parameters $\theta$, i.e., $\mathbf{v}_{w_i}$, $\mathbf{v}_{w_j}$,  are trained by maximizing the log likelihood of predicting context words in a corpus:
\begin{equation}
\label{eq:lp}
J= \sum_i^{|\mathcal{V}|}{\sum_{j \in context(i)} { \log p(w_j|w_i) }},
\end{equation}
where $context(i)$  is the set of context words of  $w_i$.

{\noindent \bf Network embedding}
Network embedding aims to map the network  $G=(V,E)$  into a low dimensional space where each vertex is represented as a low dimensional real vector. The network embedding treats the graph's vertex set $V$  as the vocabulary $\mathcal{V}$ and treats each vertex $v_i$ as a word $w_i$ in the skip-gram approach.
The corpus used for training is composed by the edge set $E$, e.g.,  in \cite{Tang2015Line},  or a set of truncated random walks from the graph, e.g., in \cite{Perozzi2014Deepwalk} and \cite{Grover2016Node2Vec}.

To train the node vectors, the objective of previous network embedding models is to predict the neighbor nodes $N(v_i)$ of a given source node $v_i$.
However, predicting a number of neighbor nodes requires modeling the joint probability of nodes, which is hard to compute.
The conditional independence assumption, i.e.,  the likelihood of observing a neighbor node is independent of observing other neighbor nodes given the source node, is often assumed \cite{Grover2016Node2Vec}.
Thus, the objective function is defined as:
\begin{equation}
\label{eq:neighbor}
J = \sum_{v_i \in V}{\log p(N(v_i)|v_i)} = \sum_{v_i \in V}{\sum_{v'_i \in N(v_i)}{\log p(v'_i|v_i)}},
\end{equation}
where $p(v'_i|v_i)$ is softmax function similar to Equation \ref{eq:softmax} except that the word vectors are replaced with node vectors.

\vspace{-0.2cm}
\section{SNE: Signed Network Embedding}
 We present our network embedding model for signed networks.
For each node's embedding, we introduce the use of both source embedding  and target embedding  to capture the two potential roles of each node.

\vspace{-0.2cm}
\subsection{Problem definition}
Formally, a signed network is defined as $G=(V, E_+, E_-)$, where $V$ is the set of vertices and $E_+$ ($E_-$) is the set of positive (negative) edges. Each edge $e \in E_+ \cup E_-$ is represented as $e_{uv}=(u,v, \varepsilon_{uv})$, where $u,v \in V$ and $\varepsilon_{uv}$ indicates the sign value of edge $e$, i.e., $\varepsilon_{uv}=1$ if  $e \in E_+$ and $\varepsilon_{uv}=-1$ if  $e \in E_-$. In the scenario of signed directed graphs,
$e_{uv}$ is a directed edge where node node $u$ denotes the source and $v$ denotes the target.
Our goal is to learn node embedding for each vertex in a signed network while capturing as much topological information as possible. For each vertex  $v_i$, its node representation is defined as $\bar{\mathbf{v}}_{v_i} = [\mathbf{v}_{v_i}:\mathbf{v'}_{v_i}]$ where $\mathbf{v}_{v_i} \in \R^{d}$ denotes its source embedding  and $\mathbf{v'}_{v_i} \in \R^{d}$  denotes its target embedding.

\vspace{-0.2cm}
\subsection{Log-bilinear model for signed network embedding}

We develop our signed network embedding by adapting the log-bilinear model such that the trained node embedding can capture node's path and sign information.  Recall that existing unsigned network embedding models are based on the skip-gram which only captures node's neighbour information and cannot deal with the edge sign.

{\bf\noindent Log-bilinear model} Given a sequence of context words $g=w_1,...,w_l$, the log-bilinear model firstly computes the predicted representation for the target word by linearly combining the feature vectors of words in the context with the position weight vectors:
\begin{equation}
\hat{\mathbf{v}}_g = \sum_{j=1}^l {\mathbf{c}_j \odot \mathbf{v}_{w_i}},
\end{equation}
where $\odot$ indicates the element-wise multiplication and $\mathbf{c}_j$ denotes the position weight vector of the context word $w_i$.
A score function is defined to measure the similarity between the predicted target word vector and its actual target word vector:
\begin{equation}
s(w_i, g) = \hat{\mathbf{v}}_g^{T} \mathbf{v}_{w_i} + b_{w_i},
\end{equation}
where $b_{w_i}$ is the bias term. The log-bilinear model then trains word embeddings $\mathbf{v}$ and position weight vectors $\mathbf{c}$ by optimizing the objective function similar to the skip-gram.

{\bf\noindent SNE algorithm} In our signed network embedding, we adopt the log-bilinear model to predict the target node based on its paths.  The objective of the log-bilinear model is to predict a target node given its predecessors along a path. Thus, the signed network embedding is defined as a maximum likelihood optimization problem. One key idea of our signed network embedding is to use signed-type vector $\mathbf{c}_+ \in \R^{d}$ ($\mathbf{c}_- \in \R^{d}$) to represent the positive (negative) edges. Formally, for a target node $v$ and a path $h=[u_1, u_2, \dots, u_l, v]$,
the model computes the predicted target embedding of node $v$ by linearly combining source embeddings ($\mathbf{v}_{u_i}$) of all source nodes along the path $h$ with the corresponding signed-type vectors ($\mathbf{c}_i$):
\begin{equation}
\label{eq:li}
\hat{\mathbf{v}}_h = \sum_{i=1}^{l} {\mathbf{c}_i \odot \mathbf{v}_{u_i}},
\end{equation}
where $\mathbf{c}_i \equiv \mathbf{c}_+$ if $\varepsilon_{u_i u_{i+1}}=1$, or $\mathbf{c}_i \equiv \mathbf{c}_-$ if $\varepsilon_{u_i u_{i+1}}=-1$ and $\odot$ denotes element-wise multiplication. The score function is to evaluate the similarity between the predicted representation $\hat{\mathbf{v}}_h$ and the actual representation $\mathbf{v'}_v$ of target node $v$:
\begin{equation}
\label{eq:score}
s(v,h) = \hat{\mathbf{v}}_h^T \mathbf{v'}_v + b_v,
\end{equation}
where $b_v$ is a bias term.

To train the node representations, we define the conditional likelihood of target node $v$ generated by a path of nodes $h$ and their edge types $q$ based on softmax function:
\begin{equation}
\label{eq:score_softmax}
p(v|h,q;\theta) = \frac{exp(s(v,h))}{\sum_{v' \in V}{exp(s(v',h))}},
\end{equation}
where $V$ is the set of vertices, and $\theta=[\mathbf{v}_{u_i},\mathbf{v'}_v,\mathbf{c},b_v]$. The objective function is to maximize the log likelihood of Equation \ref{eq:score_softmax}:
\begin{equation}
J = \sum_{v \in V} \log p(v|h,q;;\theta).
\end{equation}

\vspace{-0.2cm}
\begin{algorithm}
	\small
	\DontPrintSemicolon
	\SetKwInOut{Input}{Input}\SetKwInOut{Output}{Output}
	\Input{Signed graph $G=(V, E_+, E_-)$, embedding dimension $d$, length of path $l$, length of random walks $L$, walks per nodes $t$}
	\Output{Representation of each node $\bar{\mathbf{v}}_i = [\mathbf{v}_i:\mathbf{v'}_i]$}
	\BlankLine
	Initialization: Randomly initialize the source and target node embeddings $\mathbf{v}_i$ and $ \mathbf{v'}_i$of each node $V$

	Generate the corpus based on uniform random walk

	\For{each path $[u_1, u_2, \dots, u_l, v]$ in the corpus}{
		\For{$j=1$ to $l$}{
			\If{$\varepsilon_{u_i u_{i+1}}==1$}{
				$\mathbf{c}_i \equiv \mathbf{c}_+$
			}
			\Else{
				$\mathbf{c}_i \equiv \mathbf{c}_-$
			}
		}
		compute $\hat{\mathbf{v}}_h$ by Equation \ref{eq:li}

		compute $s(v, h)$ by Equation \ref{eq:score}

		compute $p(v|h,q;\theta)$ by Equation \ref{eq:score_softmax}

		update $\theta$ with Adagrad

	}
\caption{The SNE algorithm}
\label{alg:sne}
\end{algorithm}

Algorithm \ref{alg:sne} shows the pseudo-code of our signed network embedding. We first randomly initialize node embeddings  (Line 1) and then use random walk to generate the corpus (Line 2).
Lines 4-11 show how we specify $\mathbf{c}_i$ based on the sign of edge   $e_{u_i u_{i+1}}$. We calculate the predicted representation of the target node by combining source embeddings of  nodes  along the path with the edge type vectors (Line 12). We calculate the score function to measure the similarity between the predicted representation and the actual representation of the target node (Line 13) and compute the conditional likelihood of target node given the path (Line 14). Finally, we apply the Adagrad method  \cite{Duchi2011Adaptive} to optimize the objective function (Line 15). The procedures in Lines 4-15 repeat over each path in the corpus.

For a large network, the softmax function is expensive to compute because of the normalization term in Equation \ref{eq:score_softmax}. We adopt the sampled softmax approach \cite{Jean2014On} to reduce the computing complexity. During training, the source and target embeddings of each node are updated simultaneously. Once the model is well-trained, we get node embeddings of a signed network. We also adopt the approach  in \cite{Perozzi2014Deepwalk} to generate paths efficiently.
Given each starting node $u$, we
uniformly sample the next node from the neighbors of the last node in the path until it reaches the maximum length $L$. We then use a sliding window with size $l+1$ to slide over the sequence of nodes generated by random walk.  The first $l$ nodes in each sliding window are  treated as the sequence of path and the last node as the target node. For each node $u$, we repeat this process $t$ times.

\vspace{-0.2cm}
\section{Experiments}
To compare the performance of different network embedding approaches, we focus on the quality of their output, i.e., node embeddings. We use the generated node embeddings as input of two data mining tasks,
node classification and link prediction. For node classification, we assume each node in the network is associated with a known class label and  use node embeddings to build classifiers. In link prediction, we use node embeddings to predict whether there is a positive, negative, or no edge between two nodes. In our signed network embedding, we use the whole node representation
$\bar{\mathbf{v}}_{v_i} = [\mathbf{v}_{v_i}:\mathbf{v'}_{v_i}]$ that contains both source embedding  $\mathbf{v}_{v_i}$ and target  embedding $\mathbf{v'}_{v_i}$. This approach is denoted as $\text{SNE}_{st}$.
We also use only the source node vector $\mathbf{v}_{v_i}$  as the node representation. This approach is denoted as $\text{SNE}_{s}$.
Comparing the performance of $\text{SNE}_{st}$ and $\text{SNE}_{s}$ on both directed and undirected networks expects to help better understand the performance and applicability of our signed network embedding.

{\noindent \bf Baseline algorithms}
We compare our SNE with the following baseline algorithms.
\begin{itemize}
\item SignedLaplacian \cite{Kunegis2010Spectral}. It calculates eigenvectors of the $k$ smallest eigenvalues of the signed Laplacian matrix and treats each row vector as node embedding.
\item DeepWalk \cite{Perozzi2014Deepwalk}. It uses uniform random walk (i.e., depth-first strategy) to sample the inputs and trains the network embedding based on skip-gram.
\item LINE \cite{Tang2015Line}. It uses the breadth-first strategy to sample the inputs based on node neighbors and preserves both the first order and second order proximities in node embeddings.
\item Node2vec \cite{Grover2016Node2Vec}. It is based on skip-gram and adopts the biased random walk strategy to generate inputs. With the biased random walk, it can explore diverse neighborhoods by balancing the depth-first sampling and breath-first sampling.
\end{itemize}

{\noindent \bf Datasets}
We conduct our evaluation on two signed networks. (1) The first signed network,  \textit{WikiEditor}, is extracted from the UMD Wikipedia dataset \cite{Kumar2015Vews}. The dataset is composed by 17015 vandals and 17015 benign users who edited the Wikipedia pages from Jan 2013 to July 2014. Different from benign users, vandals edit articles in a deliberate attempt to damage Wikipedia. One edit may be reverted by bots or editors. Hence, each edit can belong to either {\em revert} or {\em no-revert} category. The WikiEditor is built based on the co-edit relations. In particular, a positive (negative) edge between users $i$ and $j$ is added if the majority of their co-edits are from the same category (different categories). We remove from our signed network those users who do not have any co-edit relations with others.
Note that in WikiEditor, each user is clearly labeled as either benign or vandal. Hence, we can run node classification task on WikiEditor in addition to link prediction. (2) The second signed network is based on the Slashdot Zoo dataset~\footnote{\url{https://snap.stanford.edu/data/}}.  The Slashdot network is signed and directed. Unfortunately,  it does not contain node label information. Thus we only conduct link prediction. Table \ref{tb:wiki} shows the  statistics of these two signed networks.

\vspace{-0.5cm}
\begin{table}[]
\centering
\caption{Statistics of WikiEditor and Slashdot}
\label{tb:wiki}
\begin{adjustbox}{max width=0.8\textwidth}
\begin{tabular}{|c|c|c|}
\hline
                   & WidiEditor              & Slashdot         \\ \hline
Type               & Undirected              & Directed         \\ \hline
\# of Users (+, -) & 21535 (7852, 13683)     & 82144 (N/A, N/A) \\ \hline
\# of Links (+, -) & 348255 (269251, 79004)  & 549202 (425072, 124130)        \\ \hline
\end{tabular}
\end{adjustbox}
\end{table}

{\noindent \bf Parameter settings}
In our SNE methods, the number of randomly sampled nodes used in the sampled softmax approach is 512. The dimension of node vectors $d$ is set to 100 for all embedding models except SignedLaplacian.  SignedLaplacian is a matrix factorization approach. We only run SignedLaplacian on WikiEditor. This is because   Slashdot is a directed graph and its Laplacian matrix is non-symmetric. As a result, the spectral decomposition  involves complex values. For WikiEditor, SignedLaplacian uses 40 leading vectors because there is a large eigengap between the 40th and 41st eigenvalues. For other parameters used in DeepWalk, LINE and Node2vec, we  use their default values based on their published source codes.

\subsection{Node classification}
We conduct node classification using the WikiEditor signed network. This task is to predict whether a user is benign or vandal in the WikiEditor signed network.
In our SNE training, the path length $l$ is 3, the maximum length of random walk path $L$ is 40, and the number of random walks starting at each node $t$ is 20.
We also run all baselines to get their node embeddings of WikiEditor. We then use the node embeddings generated by each method to train a logistic regression classifier with 10-fold cross validation.

{\noindent \bf Classification accuracy}
Table \ref{tb:class_results} shows the comparison results of each method on node classification task. Our $\text{SNE}_{st}$ achieves the best accuracy and outperforms all baselines significantly in terms of accuracy.  This indicates that our SNE  can capture the different relations among nodes by using the signed-type vectors $\mathbf{c}$. All the other embedding methods based on skip-gram  have a low accuracy, indicating they are not feasible for signed network embedding because they do not distinguish the positive edges from negative edges. Another interesting observation is that the accuracy of $\text{SNE}_{s}$  is only slightly worse than $\text{SNE}_{st}$. This is because WikiEditor is undirected.  Thus using only source embeddings in the SNE training is feasible for undirected networks.

\begin{figure*}[htbp!]
    \centering
    \begin{subfigure}[t]{0.30\textwidth}
        \centering
        \includegraphics[height=1.7in]{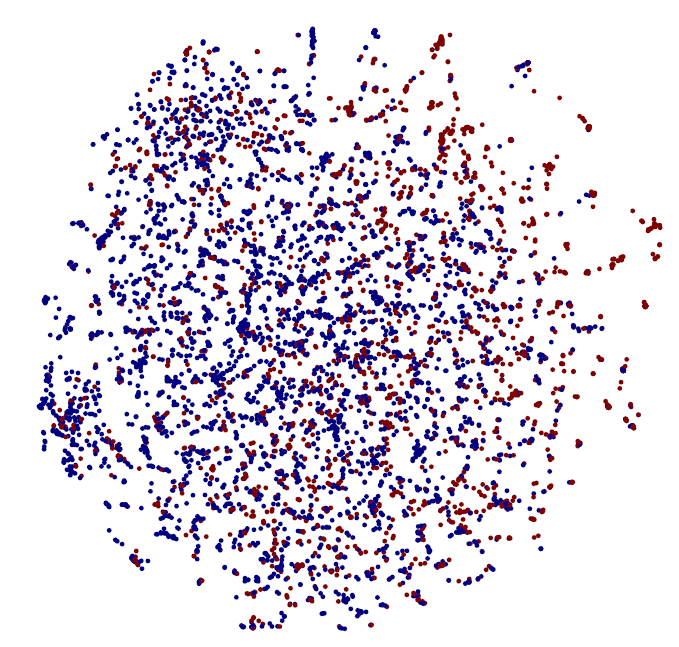}
        \caption{DeepWalk}
    \end{subfigure}%
    ~
    \begin{subfigure}[t]{0.30\textwidth}
        \centering
        \includegraphics[height=1.7in]{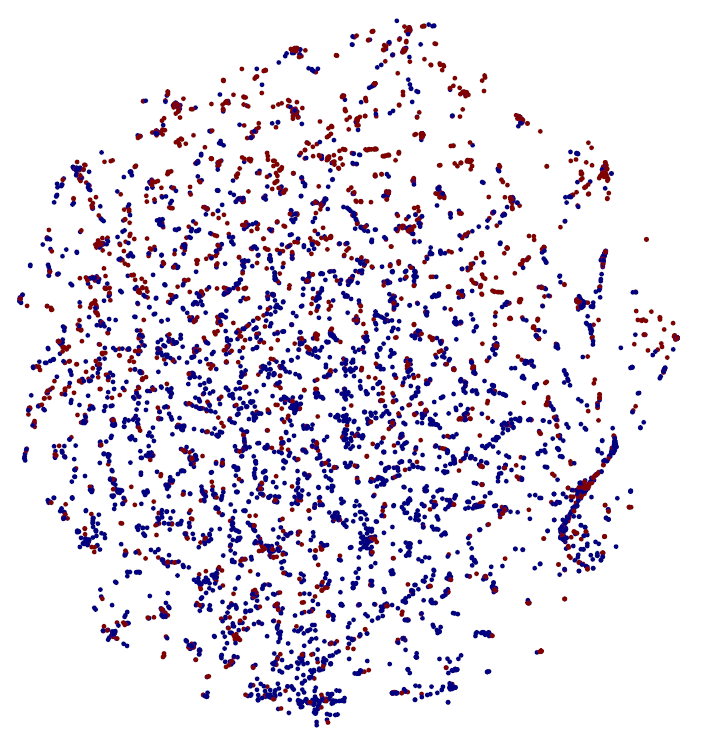}
        \caption{Node2vec}
    \end{subfigure}
    ~
    \begin{subfigure}[t]{0.30\textwidth}
        \centering
        \includegraphics[height=1.7in]{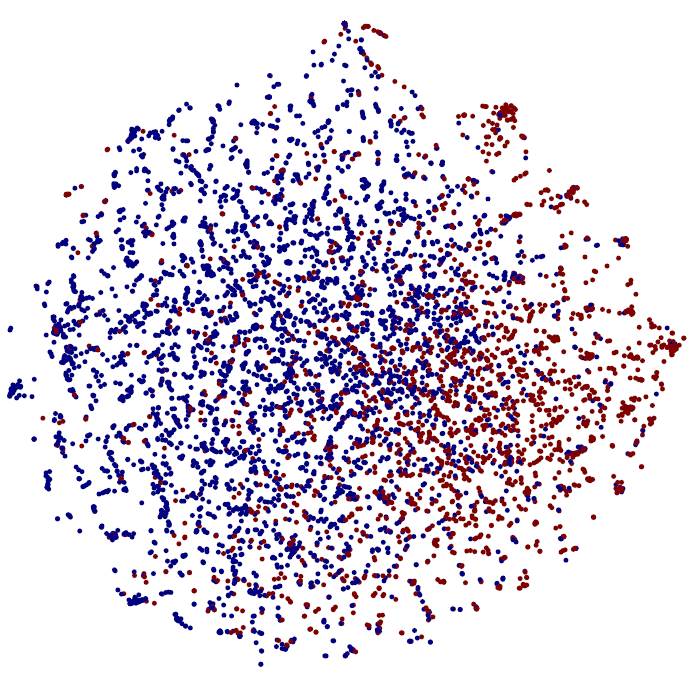}
        \caption{$\text{SNE}_{st}$}
    \end{subfigure}
    \caption{Visualization of 7000 users in WikiEditor. Color of a node indicates the type of the user. Blue: ``Vandals'', red:``Benign Users''. }
\label{fig:visual}
\end{figure*}

{\noindent \bf Visualization} To further compare the node representations trained by each approach,  we randomly choose representations of 7000 users from WikiEditor and map them to a 2-D space based on t-SNE approach \cite{Maaten2008Visualizing}. Figure \ref{fig:visual} shows the projections of node representations from DeepWalk, Node2vec, and $\text{SNE}_{st}$.
We observe that $\text{SNE}_{st}$ achieves the best and DeepWalk is the worst. Node2vec performs slightly better than DeepWalk but the two types of users still mix together in many regions of the projection space.

\vspace{-0.5cm}
\begin{table}[]
\centering
\caption{Accuracy for node classification on WikiEditor}
\label{tb:class_results}
\begin{adjustbox}{max width=0.8\textwidth}
\begin{tabular}{|c|c|c|c|c|c|c|}
\hline
         & SignedLaplacian & DeepWalk & Line    & Node2vec & $\text{SNE}_{s}$ & $\text{SNE}_{st}$ \\ \hline
Accuracy & 63.52\%         & 73.78\%  & 72.36\% & 73.85\%  & 79.63\%          & \textbf{82.07\%}  \\ \hline
\end{tabular}
\end{adjustbox}
\end{table}

\vspace{-0.5cm}

\vspace{-0.2cm}
\subsection{Link prediction}
In this section, we conduct link prediction on both WikiEditor and Slashdot signed graphs. We follow the same procedure as \cite{Grover2016Node2Vec} to make link prediction as a classification task.
We  first use  node representations to compose edge representations and then use them to build a classifier for predicting whether there is a positive, negative or no edge between two nodes.
Given a pair of nodes $(u, v)$ connected by an edge, we use an element-wise operator to combine the node vectors $\mathbf{v}_u$ and $\mathbf{v}_v$ to compose the edge vector $\mathbf{e}_{uv}$. We use the same operators as \cite{Grover2016Node2Vec} and show them in Table \ref{tb:op}.
We train and test the one-vs-rest logistic regression model with 10-fold cross validation by using the edge vectors as inputs.

\vspace{-0.8cm}
\begin{table}[]
\centering
\caption{Element-wise operators for combining node vectors to edge vectors}
\label{tb:op}
\begin{tabular}{|c|c|}
\hline
Operator   & Definition \\ \hline
Average   & $\mathbf{e}_{uv} = \frac{1}{2} (\mathbf{v}_u + \mathbf{v}_v)$  \\ \hline
Hadamard    & $\mathbf{e}_{uv} = \mathbf{v}_u * \mathbf{v}_v   $        \\ \hline
L1\_Weight & $\mathbf{e}_{uv} = |\mathbf{v}_u -  \mathbf{v}_v|  $         \\ \hline
L2\_Weight & $\mathbf{e}_{uv} = |\mathbf{v}_u -  \mathbf{v}_v|^2  $           \\ \hline
\end{tabular}
\end{table}

For Slashdot, we set the path length $l=1$ in our SNE training, which corresponds to the use of the edge list of the Slashdot graph. This is because there are few paths with length larger than 1 in Slashdot. For WikiEditor, we use the same node representations adopted in the previous node classification task.
We also compose balanced datasets for link prediction as suggested in \cite{Grover2016Node2Vec}.  We keep all the negative edges, randomly sample the same number of positive edges, and then randomly generate an equal number of fake edges connecting two nodes.  At last, we have 79004 edges for each edge type (positive, negative, and fake) in WikiEditor and  we have 124130 edges for each type in Slashdot.

{\noindent \bf Experimental results}
Table \ref{tb:lp} shows the link prediction accuracy for each approach with four different operators. We observe that our SNE with Hadamard operator achieves the highest accuracy on both WikiEditor and Slashdot.
 SNE also achieves good accuracy with the L1\_Weight and L2\_Weight. For the Average operator, we argue that it is not suitable for composing edge vectors from node vectors in signed networks although it is suitable in unsigned networks. This is because a negative edge  pushes away the two connected nodes in the vector space whereas a positive edge pulls  them together \cite{Kunegis2010Spectral}. When examining the performance of all baselines, their accuracy values are significantly lower than our SNE,  demonstrating their infeasibility for signed networks.

 We also observe that there is no big difference between $\text{SNE}_{st}$ and $\text{SNE}_{s}$ on WikiEditor whereas $\text{SNE}_{st}$ outperforms $\text{SNE}_{s}$ significantly on Slashdot. This is because WikiEditor is an undirected network and Slashdot is directed.
 This suggests it is imperative to combine both source embedding and target embedding as  node representation in signed directed graphs.

\vspace{-0.5cm}
\begin{table}[]
\centering
\caption{Comparing the accuracy for link prediction}
\label{tb:lp}
\begin{adjustbox}{max width=0.8\textwidth}
\begin{tabular}{|c|c|c|c|c|c|}
\hline
Dataset                                                                            & Approach        & Hadamard        & Average         & L1\_Weight      & L2\_Weight      \\ \hline
\multirow{6}{*}{\begin{tabular}[c]{@{}c@{}}WikiEditor\\ (Undirected)\end{tabular}} & SignedLaplacian & 0.3308          & 0.5779		 	 & 0.5465          & 0.3792          \\ \cline{2-6}
                                                                                   & DeepWalk        & 0.7744          & 0.6821          & 0.4515          & 0.4553          \\ \cline{2-6}
                                                                                   & Line            & 0.7296          & 0.6750          & 0.5205          & 0.4986          \\ \cline{2-6}
                                                                                   & Node2vec        & 0.7112          & 0.6491          & 0.6787          & 0.6809          \\ \cline{2-6}
                                                                                   & $\text{SNE}_{s}$& \textbf{0.9391} & \textbf{0.6852} & \textbf{0.8699} & 0.8775 		 \\ \cline{2-6}
                                                                                   & $\text{SNE}_{st}$& \textbf{0.9399}& 0.6043          & 0.8495          & \textbf{0.8871} \\ \hline
\multirow{5}{*}{\begin{tabular}[c]{@{}c@{}}Slashdot\\ (Directed)\end{tabular}}     & DeepWalk        & 0.6907          & \textbf{0.6986} & 0.5877          & 0.5827          \\ \cline{2-6}
                                                                                   & Line            & 0.5823          & 0.6822          & 0.6158          & 0.6087          \\ \cline{2-6}
                                                                                   & Node2vec        & 0.6560          & 0.6475          & 0.4595          & 0.4544          \\ \cline{2-6}
                                                                                   & $\text{SNE}_{s}$& 0.4789          & 0.5474          & 0.6078          & 0.6080          \\ \cline{2-6}
                                                                                   & $\text{SNE}_{st}$& \textbf{0.9328} & 0.5810          & \textbf{0.8358} & \textbf{0.8627} \\ \hline
\end{tabular}
\end{adjustbox}
\end{table}

\subsection{Parameter sensitivity}

\begin{figure}[htbp!]
    \centering
    \begin{subfigure}[t]{0.4\textwidth}
        \centering
        \includegraphics[height=1.6in,width=\linewidth]{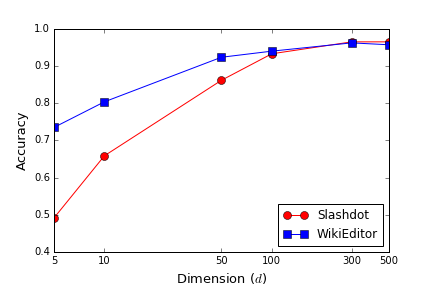}
        \caption{Dimension}
        \label{fig:dim}
    \end{subfigure}%
    \begin{subfigure}[t]{0.4\textwidth}
        \centering
        \includegraphics[height=1.6in,width=\linewidth]{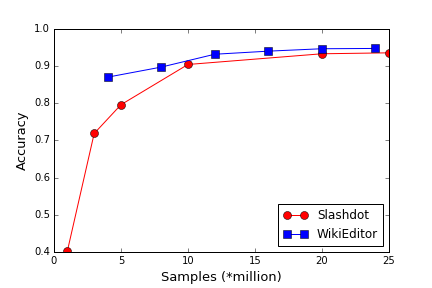}
        \caption{Sample Size}
        \label{fig:samples}
    \end{subfigure}
    \caption{The sensitivity of SNE on the WikiEditor and Slashdot}
\label{fig:perform}
\end{figure}

{\bf\noindent Vector dimension} We evaluate how the dimension size of node vectors affects the accuracy of two tasks on both WikiEditor and Slashdot. For link prediction, we use $\text{SNE}_{st}$ with Hadamard operation as it can achieve the best performance as shown in the last section. Figure \ref{fig:dim} shows how the accuracy of link prediction varies with different dimension values of  node vectors used in $\text{SNE}_{st}$ for both datasets. We can observe that the accuracy increases correspondingly for both datasets when the dimension of node vectors increases.
Meanwhile, once the accuracy reaches the top, increasing the dimensions further does not have much impact on accuracy any more.

{\bf\noindent Sample size} Figure \ref{fig:samples} shows how the accuracy of link prediction varies with the sample size used in $\text{SNE}_{st}$ for both datasets. For WikiEditor, we tune the sample size by changing  the number of random walks starting at each node ($t$). In our experiment, we set $t=5,10,15,20,25$ respectively and calculate the corresponding sample sizes. For Slashdot, we directly use the number of sampled edges in our training as the path length is one. For both datasets, the overall trend is similar. The accuracy increases with more samples. However, the accuracy becomes stable when the sample size reaches some value. Adding more samples further does not improve the accuracy significantly.

\begin{figure}[htbp!]
    \centering
    \begin{subfigure}[t]{0.4\textwidth}
        \centering
        \includegraphics[height=1.6in,width=\linewidth]{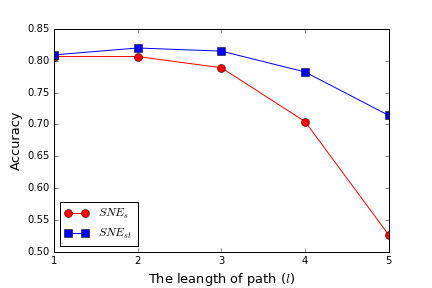}
        \caption{Node classification}
        \label{fig:node}
    \end{subfigure}%
    \begin{subfigure}[t]{0.4\textwidth}
        \centering
        \includegraphics[height=1.6in,width=\linewidth]{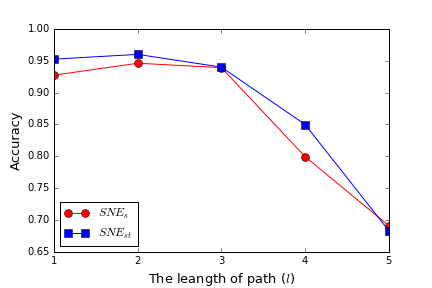}
        \caption{Link prediction}
        \label{fig:link}
    \end{subfigure}
    \caption{The sensitivity of SNE on the WikiEditor by changing the path length ($l$)}
\label{fig:path}
\end{figure}

{\bf\noindent Path length} We  use WikiEditor to evaluate how the path length $l$ affects the accuracy of both node classification and link prediction.  From Figure \ref{fig:node}, we observe that slightly increasing the path length in our  $\text{SNE}$ can improve the accuracy of node classification. This indicates that the use of long paths in our SNE training can generally capture more network structure information, which is useful for node classification. However, the performance of the $\text{SNE}_{s}$ and $\text{SNE}_{st}$ decreases when the path length becomes too large. One potential reason is that SNE uses only two signed-type vectors for all nodes along paths and
nodes in the beginning of a long path may not convey much information about the target node. In Figure \ref{fig:link}, we also observe that the accuracy of link prediction decreases when the path length increases. For link prediction, the performance depends more on local information of nodes. Hence the inclusion of one source node in the path can make our SNE learn the sufficient local information.

\vspace{-0.2cm}
\section{Related Work}
{\noindent \bf Signed network analysis} Mining signed network attracts increasing attention \cite{Leskovec2010Signed,Kunegis2010Spectral,Chiang2013Prediction,Tang2015Survey,Tang2016Node}.
The balance theory \cite{Heider1946Attitudes} and the status theory \cite{Leskovec2010Signed} have been proposed and many algorithms have been developed for tasks such as community detection, link prediction, and spectral graph analysis of signed networks  \cite{Tang2015Survey,Chiang2013Prediction,Kunegis2010Spectral,Yang2015Evaluating,DBLP:conf/icdm/WuWLL14,DBLP:conf/pakdd/WuYWLZ11,DBLP:journals/corr/LiWL16}.
Spectral graph analysis  is mainly based on matrix decomposition which is often expensive and hard to scale to large networks. It is  difficult to capture the  non-linear structure information as well as local neighborhood information  because it simply projects a global matrix to a low dimension space formed by leading eigenvectors.

{\noindent \bf Network embedding} Several network embedding methods including DeepWalk \cite{Perozzi2014Deepwalk}, LINE \cite{Tang2015Line}, Node2vec \cite{Grover2016Node2Vec}, Deep Graph Kernels \cite{Yanardag2015Deep} and  DDRW \cite{Li2016Discriminative} have been proposed. These models are based on the neural language model. Several network embedding models are based on other neural network model. For example, DNR \cite{Wang2016Structural} uses the deep auto-encoder,  DNGR \cite{Cao2016Deep} is based on a stacked denoising auto-encoder, and the work \cite{Niepert2016Learning} adopts the convolutional neural network to learn the network feature representations. Meanwhile, some works learn the network embedding by considering the node attribute information. In \cite{Yang2016Revisiting,Tu2016MaxMargin} the authors consider the node label information and present semi-supervised models to learn the network embedding.  The heterogeneous network embedding models are studied in   \cite{Chang2015Heterogeneous,Yang2015Network,Tang2015Ptea,Pan2016TriParty}. HOPE \cite{Ou2016Asymmetric} focuses on preserving the asymmetric transitivity of a directed network by approximating high-order proximity of a network. Unlike all the works described above, in this paper, we explore the signed network embedding.

\vspace{-0.2cm}
\section{Conclusion}
In this paper, we have presented SNE for signed network embedding. Our SNE adopts the log-bilinear model to combine the edge sign information and node representations of all nodes along a given path. Thus,
the learned node embeddings  capture the information of positive and negative links in signed networks.
Experimental results on node classification and link prediction showed the effectiveness of SNE.
Our SNE expects to keep the same scalability as DeepWalk or Node2vec because SNE adopts vectors to represent the sign information and uses linear operation to combine node representation and signed vectors.  In our future work, we  plan to examine how other structural information (e.g., triangles or motifs) can be preserved in signed network embedding.

\vspace{-0.2cm}
\section*{Acknowledgments}
The authors acknowledge the support from the National Natural Science Foundation of China (71571136), the 973 Program of China (2014CB340404), and the Research Project of Science and Technology Commission of Shanghai Municipality (16JC1403000, 14511108002) to Shuhan Yuan and Yang Xiang, and from National Science Foundation (1564250) to Xintao Wu. This research was conducted while Shuhan Yuan visited University of Arkansas. Yang Xiang is the corresponding author of the paper.

\end{document}